\documentclass[plb,12pt,showkeys]{revtex4-1}

\usepackage[T2A]{fontenc}

\usepackage[cp1251]{inputenc}

\usepackage[english,russian]{babel}
\usepackage{dcolumn}
\usepackage{graphicx,amsmath}

\newcommand\nn{\nonumber}

\newcommand\ba{\begin{eqnarray}}

\newcommand\ea{\end{eqnarray}}

\begin{document}
\bibliographystyle{apsrev4-1}

\title{Polarization of Radiation from Planar Channeled Positrons}

\author{S.~Bellucci}
\email{Stefano.Bellucci@lnf.infn.it}
\affiliation{INFN - Laboratori Nazionali di Frascati, 1-00044, Frascati (Rome), Italy}

\author{M.~Shatnev}
\email{mshatnev@yahoo.com}
\affiliation{NSC ``Kharkov Institute of
Physics and Technology'', 61108,  Kharkov, Ukraine}

\begin{abstract}

The Stokes parameters have been found in the framework of quantum electrodynamics for the description of polarization of radiation emitted by relativistic positrons channeled between (110) planes in Si crystal. The degree of polarization, which is simply given by the contribution of channeling radiation, has been analyzed.
Numerical calculation are presented for the frequencies that are most interesting for the sources of polarized high-energy photons.
\\

\end{abstract}

\maketitle

\selectlanguage{english}
\vskip 10mm
\begin{center}
\section{Introduction}
\end{center}

Photon beams of maximum intensity and maximum polarization are needed for performing the nuclear experiments in the intermediate energy range from $10\,MeV$ to $100\,MeV$. At the same time it is well known that in this range of energy the intensity of the radiation from $5-20\,GeV$ channeled positrons and electrons is maximum \cite{Miroshnichenko, Miroshnichenko1, Miroshnichenko2, Pivovarov, Ganenko,Uggerhoj,Uggerhoj1,Korol}. The
radiation is linearly polarized and emission occurs in a narrow
cone centered about the relativistic velocity direction with the intensity that can be an
order of magnitude larger than bremsstrahlung. Therefore it is interesting to
study the polarization of radiation under channeling and to
estimate a possibility of the creation of photon beams with high
polarization degree which will be sufficient for perfoming the nuclear experiments in the intermediate energy range, using off-axis collimation of photons under
relativistic charged particles channeling in the crystal. In addition to that, radiation emitted from a process at still higher energies-GeV-energy positrons undergoing planar channeling-has attracted special
attention in the physics community, because the model describing its production is very simple and the radiation itself is almost perfectly monochromatic.

For such radiation of soft photons one can use classical electrodynamics
to describe the emission process \cite{Akhiezer,Akhiezer1,Akhiezer2}. But when the energy of incoming electrons or positrons rises to levels above a few GeV, the classical description of
channeling radiation becomes invalid. We consider a relativistic charged particle directed into a crystal approximately parallel to one of the crystal planes.
In our case for positively charged positrons, the channels consist of
the ducts in between the crystal planes, and from the quantum mechanical point of view, the channel provides a potential well for the transverse one-dimensional particle motion, in which the particle is in a bound state. Its transitions to lower levels are accompanied by the emission of channeling radiation (CR) with frequencies related to the energy differences of the levels. The polarization properties of the radiation are completely
described with a polarization matrix $\rho_{i,j}$ which may be expressed in terms of the Stokes parameters $\xi _{n}$ \ $(n=1,2,3)$:
\begin{eqnarray} \rho_{i,j}=\frac{1}{2} \left(
\begin{array}{cc}
1+\xi_3 & \xi_1-i \xi_2 \\
\xi_1+i \xi_2 & 1 - \xi_3 \\
\end{array}
\right)_{ij} \quad i,j=x,y.
\end{eqnarray}
The degree of linear polarization is $\sqrt {\xi _{1}^{2}+\xi _{3}^{2}}$,
while the degree of circular polarization is defined by $\xi _{2}$. The angle $\beta$ defining the direction of the linear polarization in the plane perpendicular
to the direction of photon emission is determined by the following condition $\tg(2\beta)=\xi _{1}/\xi _{3}$  \cite{Baier}.

 The aim of this work is to derive the formulae for the Stokes parameters of this radiation. We obtain the formulae for the corresponding Stokes parameters, characterizing the polarization properties of the CR from
arbitrary polarized particles as the function of the set which
gives the angular dependence of the polarization
of the emitted radiation. The calculation of the CR process is
carried out by using the rules of quantum electrodynamics. The following analysis utilizes the methods used in \cite{Baier, Zhevago} and is based on the approach which was developed in \cite{Boldyshev90, Boldyshev06}. The numerical results for the CR with the energy of photons $\omega = 59\, MeV$ emitted by $14\, GeV$ positrons   channeled between (110) planes in Si are given.
\\

\section{calculation of planar CR process}

Planar channeling takes place in a one-dimensional potential i.e. a relativistic particle moving in a potential $U(x)$ periodic in
the $x$ direction, which is normal to the channeling planes, is
described by the time-independent Dirac eqution. Separation of the
wave function $\Psi $($\vec {r}$), which is the solution of the
equation, into large and small components
\begin{equation}
\label{eq1} \Psi (\vec {r}) = \left( {{\begin{array}{*{20}c}
 {\varphi (\vec {r})} \hfill \\
 {\chi (\vec {r})} \hfill \\
\end{array}} } \right)
\end{equation}

\noindent leads to a Pauli-type equation for the large components
$\varphi (\vec {r})$. Since the potential is independent of $y $
and $z$, the solution of this equation can be written in the
form
\begin{equation}
\label{eq2}
 \varphi (\vec {r}) = \sqrt {{\frac{{E + m}}{{2E}}}} \exp (i\vec {p}_{\vert
\vert}  \vec {r}_{\vert \vert}  )f(x)w{\rm ,}
\end{equation}
 where $w^{\ast} w = 1_{{\rm .}}$

This putting allows us to transform a Pauli-type equation into a
one-dimensional, relativistic Schr\"{o}dinger equation for $f$($x$)
with a relativistic particle mass
\begin{equation}
\label{eq3} \left( { - {\frac{{1}}{{2\gamma
m}}}{\frac{{d^{2}}}{{dx^{2}}}} + U(x)} \right)f_{n} (x) =
\varepsilon _{n} f_{n} (x){\rm ,}
\end{equation}
where $\gamma = E / m$\textit{} is the relativistic factor and
$\varepsilon _{n} $ is the transverse energy level of the
particle. According to quantum electrodynamics, the matrix element for CR is given by
\begin{equation}
\label{eq4}
 M_{21} =\int {\Psi _{2}^{\ast}  \vec {\alpha} \vec
{e}^{\ast} \exp ( - i\vec {k}\vec {r})\Psi _{1} d\vec {r}}=
     \int {(\varphi _{2}^{\ast}  \exp ( - i\vec
{k}\vec {r})\vec {\sigma} \vec {e}^{\ast} \chi _{1}  + \chi
_{2}^{\ast } \exp ( - i\vec {k}\vec {r})\vec {\sigma} \vec
{e}^{\ast} \varphi _{1} )d\vec {r}}  {\rm ,}
\end{equation}
\noindent where $\vec {\alpha} $ and $\vec {\sigma} $ are the
Dirac and Pauli matrices respectively, $\vec {k} $  is the photon
momentum, $\vec {e} $ is the photon polarization vector. In order to evaluate the matrix element for CR, we solve the wave equation (\ref{eq3}) and find for the absolute
square of the CR amplitude \ba
 \left\vert M_{21}\right\vert ^{2}=Cw_{1}^{\ast
}\vec{e_{1}}\left( \vec{A}^{\ast }-i\left[ \vec{B}^{\ast
}\vec{\sigma }\right] \right) w_{2} \times w_{2}^{\ast
}\vec{e_{2}}^{\ast }\left( \vec{A}^{\ast }+i\left[ \vec{B}^{\ast
}\vec{\sigma }\right] \right) w_{1}
 \ea
where $C=\left( 2\pi \right) ^{4}\frac{\left( E+m\right) \left(
E'+m\right) }{4EE'}\delta ^{2}\left( \vec{p_{\Vert
}}-\vec{p}'-\vec{k_{\Vert }}\right) $, and $\vec{A},\ \vec{B},\
I_{1,\ }I_{2\ }$ are giving by the following expressions
\begin{equation}
\label{eq6}
\begin{array}{l}
 A_{x} = 2I_{2} (1 + {\frac{{\omega} }{{2{E}'}}}),\quad A_{y} = 0,\quad
  A_{z} = 2I_{1} (1 + {\frac{{\omega} }{{2{E}'}}}),   \\
 B_{x} = {\frac{{\omega} }{{E}}}(\theta \cdot I_{1} {\kern 1pt} \cos \varphi
- I_{2} ),\quad B_{y} = {\frac{{\omega} }{{{E}'}}}\theta \cdot I_{1}
(1 + {\frac{{\omega} }{{E}}})\sin \varphi ,\quad B_{z} = {\frac{{\omega
}}{{{E}'}}} \cdot {\frac{{m}}{{E}}} \cdot I_{1},
\\
 I_{1} = \int {\exp ( - ik_{x} x) \cdot f_{2}^{\ast}  (x)} \cdot f_{1}
(x)dx,\quad   I_{2} = - {\frac{{i}}{{E}}}\int {\exp ( -
ik_{x}}  x) \cdot f_{2}^{\ast}  (x){\frac{{df_{1} (x)}}{{dx}}}dx.
\end{array}
\end{equation}
\\
\section{The Stokes parameters for channeling radiation}
In our analysis we use the condition $\omega \ll E$, which is
correct for this CR case. We also do not take into account here the
interaction between the particle`s spin and the potential of
planes. For the description of polarization, we use here the set of
vectors \{$\vec {e}_{1} ,\;\vec {e}_{2} ,\;\vec {n}$\}, which are
used in \cite{Baier} and we give it in terms of the vectors $\vec {p}_{1} = \vec {p} - \vec
{n}(\vec {n} \cdot \vec {p}),\quad {\vec {p}}'_{1} = {\vec {p}}' -
\vec {n}(\vec {n} \cdot {\vec {p}}')\quad (\vec {p},\;{\vec {p}}'$
are the momentum of the particle before and after radiation
respectively, and $\vec {n}$ is the direction in which photon is
emitted ) in the next form
\begin{equation}
\label{eq4} \vec{e_{1}}=\frac{\vec{p_{1}}}{\left\vert
\vec{p_{1}}\right\vert },\
\vec{e_{2}}=\frac{\vec{p_{1}}^{2}\vec{p_{1}}'-\vec{p_{1}}\left(
\vec{p_{1}}\cdot \vec{p_{1}}'\right) }{\left\vert
\vec{p_{1}}\right\vert
\sqrt{\vec{p_{1}}^{2}\vec{p_{1}}'^{2}-\left( \vec{p_{1}}\cdot
\vec{p_{1}}'\right) ^{2}}}.
\end{equation}
The set of vectors \{$\vec {e}_{1} ,\;\vec {e}_{2} ,\;\vec {n}$\}
forms an orthogonal basis and the vector $\vec {e}_{1} $ lies in the
radiation plane $(\vec {k},\vec {p})$. These vectors are related
with $\theta ,\;\varphi $ (spherical coordinates of the system in
which the spectrum and angular characteristics are calculated;
here a relativistic particle moving along $z$ - direction and azimuthal
angle $\varphi $ is counted out from $x$ - direction, which is normal
to the channeling planes; $\theta \ll 1;\;$) by the next relations
\ba
 \label{eq5} \vec {e}_{1} = (- \cos \varphi ,\; - \sin \varphi
,\;\theta \;), \quad \vec {e}_{2} = (\;\sin \varphi ,\; -
\cos \varphi ,\;0\;), \quad \bar  {n} = (\;\theta \cos \varphi
,\;\theta \sin \varphi ,\;\cos \theta \;).
 \ea
An arbitrary vector $\vec {R}$=$(R_{x} ,R_{y} ,R_{z} )$ in the
coordinate system ${\left\{ {\vec {e}_{1} ,\vec {e}_{2} ,\vec {n}}
\right\}}$ is written in the form
\ba
 R_{1} \approx - R_{x} \cos \varphi - R_{y} \sin
\varphi + R_{z} \theta ,{\kern 1pt} \; \quad R_{2} \approx R_{x}
\sin \varphi - R_{y} \cos \varphi ,\; \nn \\ R_{3} \approx R_{x}
\theta \cos \varphi + R_{y} \theta \sin \varphi + R_{z} {\rm .}
 \ea
Introducing the density matrices for relativistic particle and
photon and after corresponding calculations we find the next
general expressions of the Stokes parameters for the outgoing
photon
\begin{align}
\label{eq110}
 \xi _{1}& = (8 / \Sigma L^{e})(1 + \omega / {E}')(\theta \cdot
ReI_{1} \cdot I_{2}^{\ast}-{\left| {I_{2}}
\right|}^{2}\cos \varphi ) \cdot \sin \varphi ,  \nn \\\noindent
\xi _{2} &= \varsigma _{3}\left( 8/\Sigma L^{e}\right) \left(
2\omega /E'\right) \left( 1+2\omega /E'\right) 2\theta \times \textbf{Re}(I_{1}^{\ast }I_{2})\cos \varphi \nn\\&
-\varsigma _{3}\left( 8/\Sigma L^{e}\right) \left( 2\omega
/E'\right) \left( 1+2\omega /E'\right) \left\vert I_{1}\right\vert
^{2}\theta ^{2}-\varsigma _{3}\left( 8/\Sigma
L^{e}\right) \left( 2\omega /E'\right) \left( 1+2\omega /E'\right)
\left\vert I_{2}\right\vert ^{2}- \nn\\& -\varsigma _{3}\left(
8/\Sigma L^{e}\right) \left( \omega ^{2}m^{2}/{E'}^{2}E^{2}\right)
\left\vert I_{1}\right\vert ^{2}-\varsigma _{2}\left(
8/\Sigma L^{e}\right) \left( 2\omega
m/E'E\right) \textbf{Re}(I_{1}^{\ast }I_{2})\sin \varphi -\nn\\
&-\varsigma _{1}\left( 8/\Sigma L^{e}\right) \left( 2\omega
m/E'E\right) \left\vert I_{1}\right\vert ^{2}\theta + \varsigma _{1}\left( 8/\Sigma L^{e}\right) \left( 2\omega
m/E'E\right) \textbf{Re}(I_{1}^{\ast }I_{2})\cos \varphi,\\\xi
_{3} &=\left( 4/\Sigma L^{e}\right) \left( 1+2\omega /E'\right)
\left\vert I_{1}\right\vert ^{2}\theta ^{2}+ \left(
4/\Sigma L^{e}\right) \left( 1+2\omega /E'\right) \left\vert
I_{1}\right\vert ^{2}\cos 2\varphi -  \nn\\ &-\left( 4/\Sigma
L^{e}\right) \left( 1+2\omega /E'\right) 2\theta
\textbf{Re}(I_{1}^{\ast }I_{2})\cos \varphi,\nn
\end{align}
where $E$ and ${E}' = E - \omega $ are the energy of relativistic
particle before and after radiation respectively, $m$ is the
electron mass, $\omega $ is the energy of emitted photon, $\zeta
_{1} ,\zeta _{2} ,\zeta _{3} $ are the components of the unit spin
vector $\vec {\varsigma} $ of the initial particle given in the
coordinate system ${\left\{ {\vec {e}_{1} ,\vec {e}_{2} ,\vec {n}}
\right\}}$. A normalisation factor $\Sigma L^{e}$ is found by the
formula
\begin{equation}
\label{eq10}
\begin{array}{l}
\Sigma L^{e}=4\left( 1+\omega /E'+\omega ^{2}/2{E'}^{2}\right)
\theta ^{2}\left\vert I_{1}\right\vert ^{2}+4\left( 1+\omega
/E'+\omega ^{2}/2{E'}^{2}\right) \left\vert I_{2}\right\vert
^{2}-\\-8\left( 1+\omega /E'+\omega ^{2}/2{E'}^{2}\right) \theta
\textbf{Re}(I_{1}\times I_{2}^{\ast })\cos \varphi +\left(
\omega ^{2}m^{2}/2{E'}^{2}E^{2}\right) \left\vert I_{1}\right\vert
^{2}
\end{array}
\end{equation}
We can obtain from the conservation laws of energy and momentum the
next relation between the direction of $\vec {k}$, the difference
of transverse energies before and after radiation $(\varepsilon
_{n} - \varepsilon _{{n}'} )$, and the radiation
frequency $\omega$:
\begin{equation}
\label{eq12} \theta ^{2} = {\frac{{2E(E - \omega )(\varepsilon
_{n} - \varepsilon _{{n}'} ) - m^{2}\omega} }{{E\omega (E - \omega
\cos ^{2}\varphi )}}}{\rm .}
\end{equation}
It may be shown, that in the cases of photons with energy $\omega
\ll E$ there is the following relation for integrals $I_{1} {\kern
1pt} ,I_{2} $:
\begin{equation}
\label{eq13} I_{2} = I_{1} ({\frac{{\varepsilon _{n} - \varepsilon
_{{n}'}} }{{k_{x}} }} + {\frac{{k_{x}} }{{2E}}}){\rm .}
\end{equation}
Then, using (\ref{eq12}), the last expression may be written in the
form
\begin{equation}
\label{eq14} I_{2} = I_{1} {\frac{{\gamma ^{ - 2} + \theta ^{2}(1
- {\frac{{\omega ^{2}}}{{E^{2}}}}\cos ^{2}\varphi )}}{{2\theta (1
- \omega / E)\cos \varphi }}}{\rm .}
\end{equation}
This allows us to except $I_{1} {\kern 1pt} ,I_{2} $ from
(\ref{eq10}), and the polarization of CR is independent
on the planar-continuum potential.
\\
\section{Conclusion}
The degree of circular polarization $\xi _{2}$  of  Eq.(\ref{eq110}) shows that only polarized positrons produce circularly polarized CR and it is proportional to $\omega / E$. In general,
the circular polarization from longitudinally polarized particle
is considerably greater than from transversely polarized particle
exactly in the same way as it occurs in the case of bremsstrahlung.
The linear polarization of CR is not dependent upon the positron
spin  as it follows from (\ref{eq110}), and its degree is given by
\begin{equation}
\label{eq15} P = \sqrt {\xi _{1}^{2} + \xi _{3}^{2}}  {\rm .}
\end{equation}
If we substitute into (\ref{eq110}) the expressions (\ref{eq10}) for $\Sigma L^{e}$ and (\ref{eq14}) for $I_{2}$ , we can obtain final analytical expressions for the Stokes parameters, and taking (\ref{eq15}) into consideration,
we can find the degree of linear polarization. These rather complicated formulae are not given here.
The results of the numerical calculations under these formulae for $E
= 14 \, GeV$ and $\omega = 59 \, MeV$ are presented in figs. 1 and 2. Fig.1 shows calculated degree of linear polarization
$P\left( {\theta ,\,\varphi}  \right)$ of CR as a function of the polar angle $\theta$ and the azimuthal angle $\varphi$, with the energy of photons $\omega = 59\,MeV$ from $14\,GeV$  positrons   channeled between (110) planes in silicon. Note that in our calculations $0 \leq \theta \leq 4\gamma^ {-1} \ll 1$  at all azimuths i.e. the photons are emitted in the forward direction.
As can be seen from Fig.1,  the degree of linear polarization of planar CR is
almost completely linearly polarized in the direction normal to the crystal planes. Fig.2 illustrates calculated degree of circular polarization $\xi _{2} \left( {\theta ,\,\varphi} \right)$, which is 0.426\% $\div $ 0.429\%, for various values of $\theta$ and $\varphi$, when channeled positrons are completely polarized antiparallel to the direction of the original beam $(\zeta _{3} =-1)$.

Thus, we have shown that the channeling radiation emitted in the forward direction by planar channeled positrons is completely linearly polarized. It is important for practical applications of channeling radiation in photonuclear physics, and developed computer simulation method will allow one to get the CR beam parameters which are needed for performing the nuclear experiments in the intermediate energy range from $10\,MeV$ to $100\,MeV$.
\begin{figure}[t]
\begin{center}
\includegraphics{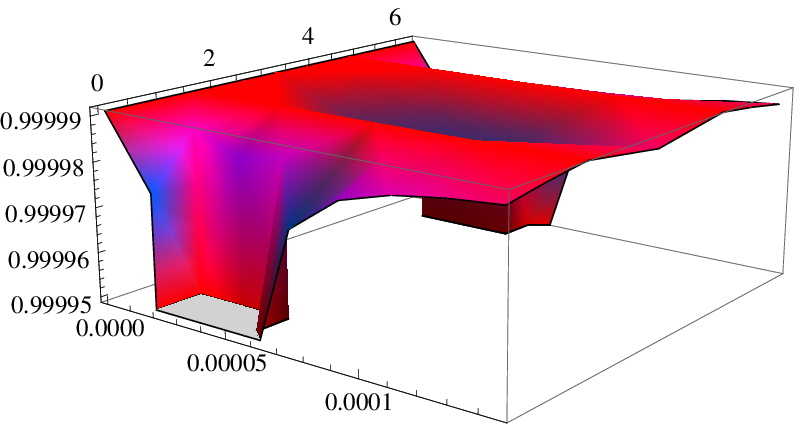}
\emph{\textbf{Fig.\;1.}}\, {\emph{Calculated degree of linear polarization
$P\left( {\theta ,\,\varphi}  \right)$ of CR as a function of the polar angle $\theta$ and the azimuthal angle $\varphi$, with the energy of photons $\omega = 59\,MeV$ from $14\,GeV$  positrons channeling in Si (110), $T = 293^{ \circ} \,K$.
}}
\end{center}
\end{figure}
\begin{figure}[t]
\begin{center}
\includegraphics{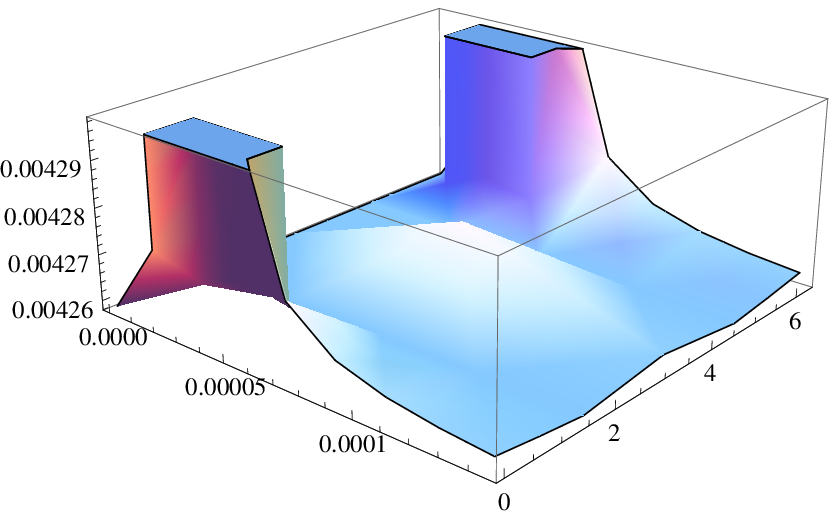}
\emph{\textbf{Fig.\;2.}}\, {\emph{Calculated degree of circular polarization
$\xi _{2} \left( {\theta ,\,\varphi} \right)$ of CR as a function of the polar angle $\theta$ and the azimuthal angle $\varphi$, with the energy of photons $\omega = 59\,MeV$ from $14\,GeV$  positrons channeling in Si (110), $T = 293^{ \circ} \,K$, $(\zeta _{1} = \zeta
_{2} = 0,\,\zeta _{3} =-1)$.
}}
\end{center}
\end{figure}
\\
\def\refname{References and Notes}

\end{document}